\documentclass[namedreferences]{SolarPhysics}
\usepackage[optionalrh]{spr-sola-addons} 
\usepackage{graphicx}        
\usepackage{color}           
\usepackage{url}             

\newcommand{\aap}{    {\it Astron. Astrophys. }}

\newcommand{\aapr}{   {\it Astron. Astrophys. Rev. }}

\newcommand{\apj}{    {\it Astrophys. J. }}
\newcommand{\apjl}{   {\it Astrophys. J. Lett. }}

\newcommand{\lrsp}{  {\it Living Rev. Sol. Phys. }}
\newcommand{\mnras}{  {\it Mon. Not. Roy. Astron. Soc. }}

\newcommand{\solphys}{{\it Solar Phys. }}

\newcommand{\ssr}{    {\it Space Sci. Rev. }}

\begin{document}
\begin{article}
\begin{opening}
\title{Statistical Properties of Photospheric Magnetic Elements Observed by SDO/HMI}

\author{M.~ \surname{Javaherian}$^1$\sep
        H.~\surname{Safari}$^1$\sep
        N.~\surname{Dadashi}$^{1}$
        M.~J.~\surname{Aschwanden}$^{2}$
        }
\runningauthor{Javaherian \textit{et al.}}
\runningtitle{%
Statistics of Solar Magnetic Elements}
\institute{
$^{1}$ Department of Physics, University of Zanjan, University Blvd., 45371-38791, Zanjan, I. R. Iran,
~email: \url{safari@znu.ac.ir} \\
$^{2}$ Lockheed Martin, Solar and Astrophysics Laboratory, Org. A021S, Bldg. 252, 3251 Hanover St., Palo Alto, CA 94304, USA
}
\begin{abstract}
Magnetic elements of the solar surface are studied (using the 6173 \AA ~Fe~{\small
\small I} line) in magnetograms recorded with the high$-$resolution
\textit{Solar Dynamics Observatory}$/$\textit{Helioseismic and Magnetic Imager}
(SDO$/$HMI). To extract some statistical and physical properties of these
elements (\textit{e.g.}, filling factors, magnetic flux, size, lifetimes),
the {\sl Yet Another Feature Tracking Algorithm (YAFTA)}, a region-based method,
is employed.  An area with 400$^{\prime\prime}\times$400$^{\prime\prime}$ was
selected to investigate the magnetic characteristics during the year 2011.
The correlation coefficient between filling factors of negative and positive
polarities is 0.51. A broken power law fit was applied to the frequency
distribution of size and flux. Exponents of the power$-$law distributions
for sizes smaller and greater than 16 arcsec$^{2}$  were found to be
-2.24 and -4.04, respectively. The exponents of power$-$law distributions
for fluxes smaller and greater than 2.63$\times$10$^{19}$ Mx were found to be
-2.11 and -2.51, respectively. The relationship between the size ($S$) and
flux ($F$) of elements can be expressed by a power$-$law behavior in the form
of $S\propto F~^{0.69}$. The lifetime and its relationship with the flux and
size of quiet-Sun (QS) elements are studied during three days. The code
detected patches with lifetimes of about 15 hours, which we call
long-duration events. It is found that more than 95\% of the magnetic elements
have lifetimes of less than 100 minutes. About 0.05\% of the elements were found
with lifetimes of more than 6 hours. The relationships between the
size ($S$), lifetime ($T$), and the flux ($F$) for
patches in the QS, indicate the power$-$law relationships $S\propto T~^{0.25}$
and $F\propto T~^{0.38}$, respectively. Executing a detrended fluctuation
analysis of the time series of new emerged magnetic elements, we find a Hurst
exponent of 0.82, which implies long-range temporal correlation in the system.
\end{abstract}
\end{opening}

\section{Introduction}

Undeniably, most of the solar atmospheric phenomena arise from photospheric magnetic
fields, consisting of a wide range of sizes and strengths. The smallest magnetic
features, with positive and negative polarities, known as ``patches'', are well
observed in magnetograms, which ubiquitously cover the solar surface with a
scale-free distribution \cite{Parnell}. The field strength of the smallest
elements can reach up to 1--2 kG, including sizes ranging from $10^{14}$ to
$10^{16}$ cm$^2$ \cite{Stenflo}, with a typical flux of $10^{17}$--$10^{19}$ Mx
(Hagenaar \textit{et al.}, 2003; Solanki, Inhester, and Sch$\ddot{u}$ssler, 2006; Priest, 2014).
Fragments with a larger flux area are affected by four dominant processes --
emergence, coalescence, fragmentation and cancellation (Close \textit{et al.},
2005; DeForest \textit{et al.}, 2007), which can affect the event occurrence
in the upper layers of the solar atmosphere. Studying their evolution,
physical properties, and statistics, may give us a better understanding of the
underlying physical processes.

Magnetic fields can appear as network regions elongated in intergranular lanes with
an average of order 1 kG, or internetwork regions over the cell interiors in the
quiet-Sun (QS), with an average of order 1 hG (Solanki, 1993; De Wijn \textit{et al.},
2009; Go\v{s}i\'{c} \textit{et al.}, 2014). Some of the statistical properties of
magnetic elements were reported by  Go\v{s}i\'{c} \textit{et al.}, (2014, 2016),
which quantify the physical parameters of magnetic features in the photosphere.
Details on the magnetic field evolution in the solar atmosphere and photosphere
can be found in Rieutord and Rincon (2010), Stein (2012), and
Wiegelmann \textit{et al.} (2014).

Because of the influence of the surface magnetic field on other events and its
important role in energy release mechanisms, the relation between photospheric
magnetic fields and other solar features (\textit{e.g.}, flares, and CMEs)
were investigated in many articles (Schrijver and Title, 2003; Close \textit{et al.},
2005; Wang \textit{et al.}, 2012; Gosain, 2012; Burtseva and Petrie, 2013;
Honarbakhsh \textit{et al.}, 2016). On the other hand, some work paid attention
to the statistical investigation of magnetic patches. A distribution of magnetic
fluxes and areas relating to active regions and QS with different kinds of
results were discussed in Schrijver \textit{et al.} (1997a), Schrijver
\textit{et al.} (1997b), Abramenko \& Longcope (2005), and Hagenaar
\textit{et al.} (2003). One of the latest work that studies magnetic
fields in both active regions and QS for more than five decades, investigated the
functional form of magnetic flux distribution based on a \textit{clumping} algorithm \cite{Parnell}. They found that the flux of all
features follow a power$-$law distribution with an exponent of approximately -1.85.
Lamb \textit{et al.} (2013) tracked 2$\times$10$^{4}$ features in the
QS to survey their birth (and death) during processes like cancellation, emergence,
and fragmentation within a lifetime of an hour.

To analyze photospheric magnetic fields, solar space missions such as
the \textit{Solar and Heliospheric Observatory} (SOHO) and the \textit{Solar Dynamics Observatory} (SDO) with their high spatial and temporal resolution, \textit{i.e.}, \textit{Michelson Doppler Imager} [MDI; Scherrer \textit{et al.}, 1995] and \textit{Helioseismic and Magnetic Imager} [HMI; Schou \textit{et al.}, 2012],
respectively, provide a vast amount of information in the form of
magnetogram images. To optimize the statistical data analysis techniques, it is necessary to develop and use automatic detection software (\textit{e.g.}, see Aschwanden, 2010; Javaherian \textit{et al.}, 2014; Arish \textit{et al.}, 2016). One of the robust and applicable algorithms is the code
\textit{Yet Another Feature Tracking Algorithm} (YAFTA) which is developed
to segment and track small- and large-scale magnetic areas in magnetograms
\cite{DeForest}.

In a first step, in order to investigate magnetic patches in a wide range of
scales, we employed the code YAFTA to segment (not to track) and extract the
physical parameters (\textit{e.g.}, size distribution, filling factors,
magnetic flux) of both negative and positive polarities within a large
area in the magnetograms, taken from SDO$/$HMI during the year 2011.
In addition to the flux distribution of magnetic elements, which has been
discussed in Parnell \textit{et al.} (2009), we focus on the other statistical
parameters of patches, such as size and lifetime frequencies. Moreover, to
find scaling laws between size, lifetime, and flux for magnetic features,
we cropped smaller sizes of magnetograms in the QS with a much finer cadence
of 45 s for a duration of three days.

The layout of this paper is as follows: the description of data sets is
presented in Section \ref{Data}. The results are presented and discussed
in Section \ref{Res}. Concluding remarks are given in Section \ref{Conc}.

\section{Description of Data Sets}\label{Data}

With the \textit{Helioseismic and Magnetic Imager} (HMI), full solar disk images
were observed in the Fe~{\small \small I} absorption line at 6173 \AA , with a
resolution of 0.50$\pm$0.01 arcsec (equivalent to $3.58 \times 10^7$ cm on
the solar surface). Level-1 data are corrected for
exposure time, dark current, gain, flat field, and cosmic-ray hits.
We applied the code \textit{Yet Another Feature Tracking Algorithm}
(YAFTA) on the dataset recorded during the year 2011, from January 1
until December 31, taken at 13:00 UT with a cadence of one image per day.

\begin{figure} 
\centerline{\includegraphics[width=1.3\textwidth,clip=]{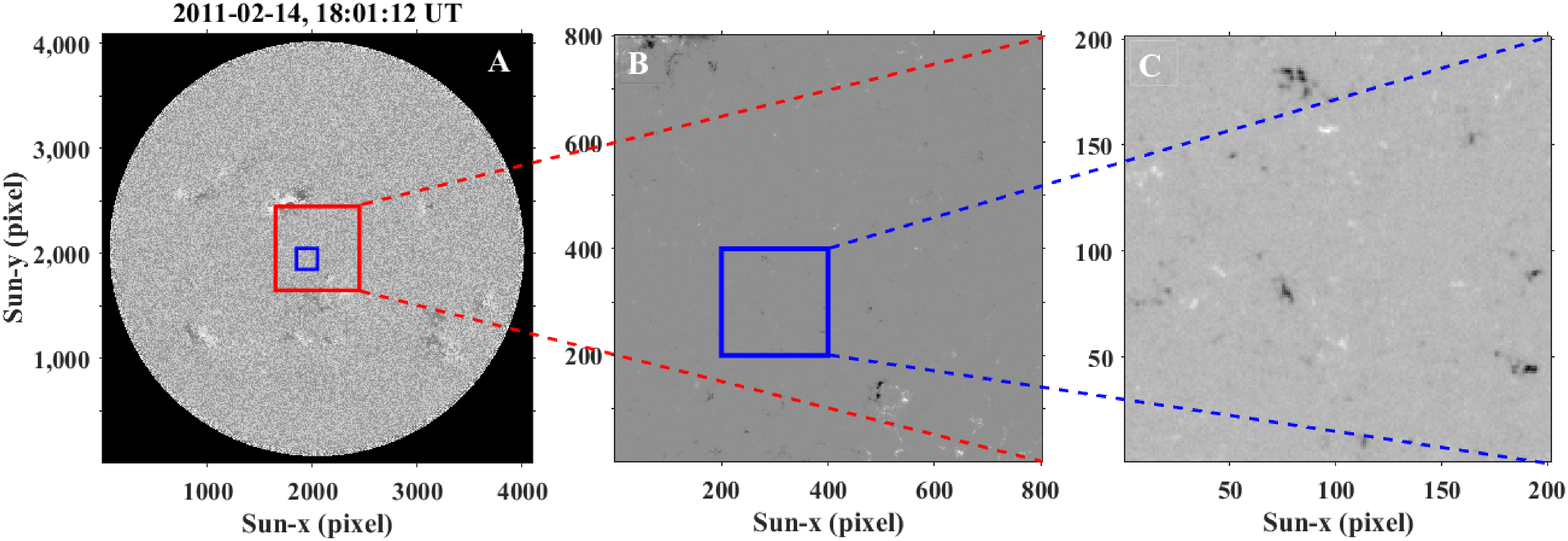}}
\caption[]{The SDO$/$HMI full-disk magnetogram of the Sun recorded on 14 February 2011 (18:01:12 UT)(A). The cutout image with area of $400\times400$ arcsec$^2$ (B) from the solar equatorial region is selected to extract physical parameters of positive and negative elements (such as size distributions, flux distributions, and filling factors) during the year 2011 with a cadence of one image per day. An image tile with area of $100\times100$ arcsec$^2$ (C) is cropped from all images to investigate physical parameters of the magnetic elements especially their lifetimes for three days (14$-$16 February 2011) in the QS with a time lag of 45 s between frames.} \label{fig1}
\end{figure}

The square area with a size of 400$^{\prime\prime}
\times$400$^{\prime\prime}$ at disk center is selected to ignore
projection effects (\textit{e.g.}, McAteer \textit{et al.}, 2005;
Alipour and Safari, 2015). This area is shown
in Figure \ref{fig1}B (red box in Figure \ref{fig1}A). Both the size
and flux distributions, the daily magnetic flux, and filling factors
of polarities are extracted. In the next step, we focus on lifetimes,
flux, and size in the quiet-Sun (QS). We downloaded level-1 HMI images
for three days from 14 February to 16 February 2011 with a cadence of
45 s. Within 5750 consecutive frames, the area with the size of
100$^{\prime\prime} \times$100$^{\prime\prime}$ is extracted from the
solar equatorial region along the line of sight. This area is shown
in Figure \ref{fig1}C (blue boxes in Figure \ref{fig1}A, and \ref{fig1}B).

The de-rotation procedure (drot\_map.pro), available in the SSW/IDL
package, is applied to coalign the images in the sequence. A subsonic
filter was applied on our data cubes to remove the global solar p-mode
oscillations for features with horizontal speeds faster than the
7 km s$^{-1}$ sound speed.

\section{Results and Discussion}\label{Res}

One of the robust software codes is \textit{Yet Another Feature Tracking
Algorithm} (YAFTA), which is improved by Deforest \textit{et al.} (2007)
and was employed in a number of studies (\textit{e.g.}, Burtseva
and Petrie, 2013; Stangalini \textit{et al.}, 2013; Go\v{s}i\'{c}
\textit{et al.}, 2014). This software is accessible from the
IDL library and computes properties of elements (magnetic flux with
sign of polarity, size, etc.) and tracks them in a consecutive list
of images. This study investigates statistical characteristics of
magnetic elements during the year 2011 within a large area
(Figure \ref{fig1}B), with a cadence of one image per day,
covering 3 days (14--16 February) with a smaller area in the QS
(Figure \ref{fig1}C), with a time step of 45 s between frames.

\subsection{Statistical properties of magnetic elements during the year 2011}

There are three kinds of algorithms for identifying structures of flux concentrations. In the \textit{clumping} algorithm, pixels with fluxes higher than the determined threshold are considered as single feature \cite{Parnell0}. In the \textit{downhill} approach, after thresholding, one feature per local maximum region is assumed to be one identified magnetic element \cite{Welsch}. In this method, the number of detected patches is more than that of obtained by the \textit{clumping} method. \textit{Curvature} algorithm determines the boundaries as convex core around each local flux maximum \cite{Hagenaar0}. In this method, the size of features decreases in comparison with the two previous approaches. It seems that \textit{downhill} and \textit{clumping} techniques produce better segmentation results than the \textit{curvature} method \cite{DeForest}.

We selected an area with $1.5\times1.5$ arcsec$^2$ ($3\times3$ pixels$^2$),
close to the value obtained by Parnell \textit{et al.} (2009),
as a minimum threshold in the grouping of pixels by the YAFTA's
\textit{downhill} algorithm. The threshold of the magnetic field
is selected to be 25 G (\textit{i.e.}, the code ignores pixels below
this threshold in magnetograms). As mentioned in the previous section,
to avoid projection effects, the area with a size of $400\times400$
arcsec$^2$ is selected from solar equatorial regions (Figure \ref{fig1}A,
red box). During the year 2011, the code detected 201,737 and 227,056
magnetic elements with positive and negative polarities, respectively.

\begin{figure} 
\centerline{\includegraphics[width=1.3\textwidth,clip=]{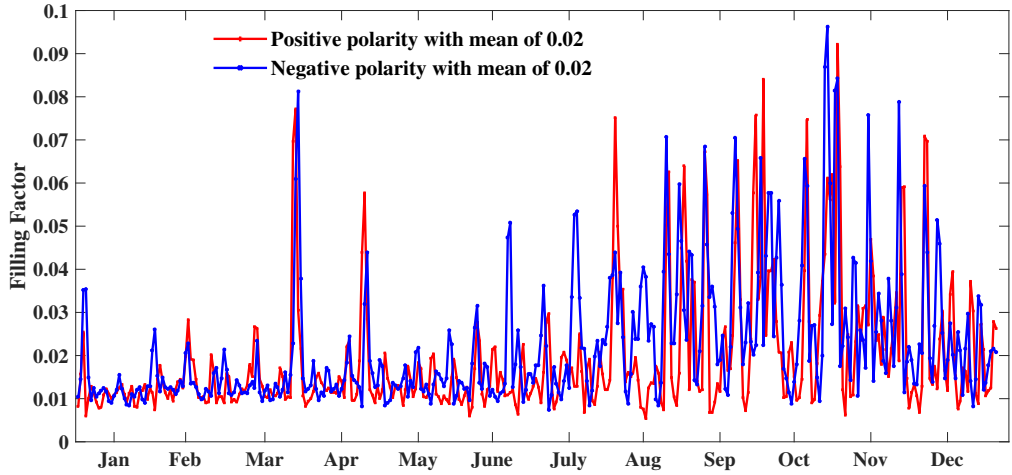}}
\caption[]{Filling factors (area) of positive (red) and negative (blue) polarities (elements) are plotted from January 1 until December 31, 2011. The \textit{Pearson} and \textit{Spearman} correlations between these time series are 0.51 and 0.29, respectively.} \label{fig2}
\end{figure}

\begin{figure} 
\centerline{\includegraphics[width=1.3\textwidth,clip=]{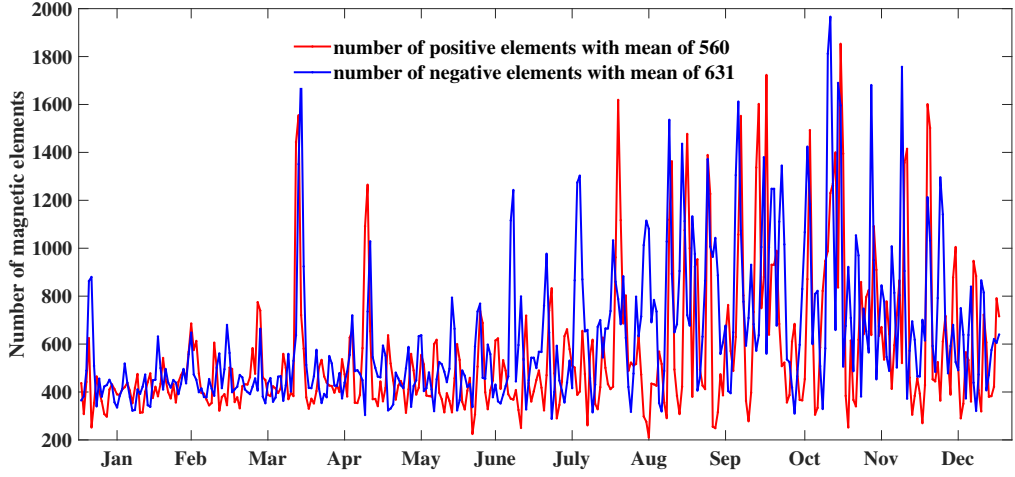}}
\caption[]{Daily number of positive (red) and negative (blue) elements are shown from January 1 until December 31, 2011.} \label{fig3}
\end{figure}

The area filling factors of both positive and negative elements with
their mean values are shown in Figure \ref{fig2}. The \textit{Pearson}
and \textit{Spearman} correlation coefficients between filling factors
of both time series were obtained with values of 0.51 and 0.29, respectively.
The daily magnetic number of the two magnetic polarities is displayed in
Figure \ref{fig3}. The size-frequency distribution of the magnetic
elements is extracted (Figure \ref{fig4}). In order to minimize the
difference between the histogram and an unknown density function,
a data-based algorithm is used to find the optimum bin number in all
histograms \cite{Shimazaki}. We fit a broken power law function to
the frequency distribution of sizes. For sizes smaller than 16 arcsec$^2$,
the exponent was obtained to be -2.24, and for the greater sizes,
an exponent of -4.04. Moreover, the power law fit was applied to
data points using the maximum likelihood estimator method [MLE;
Clauset, Shalizi, and Newman, 2009]. Using MLE, the exponent of the fit
for the whole range of data points yields the value of -2.71 $\pm$0.16.
The maximum area of patches with negative polarity was found to have
a size of 377 arcsec$^2$ in this analysis. The average of daily
negative and positive flux with a yearly mean value of
2 $\times$ 10$^{18}$ Mx is plotted in Figure \ref{fig5}.

The flux-frequency distribution of patches is plotted in Figure \ref{fig7}.
A broken power law fit was applied to fluxes smaller, and greater than
2.63$\times$10$^{19}$ Mx, yielding slope values of -2.11 and -2.51,
respectively. Besides, by applying the MLE method on the whole range
of the power law distribution, a power exponent of -2.15 $\pm$0.15
is obtained, which is close to the values in Parnell \textit{et al.} (2009).
The peak value of the flux is 1.52$\times$10$^{21}$ Mx for
the positive polarity group.

The scatter plots of the patch size versus flux, along with the mean values
(red points) for each bin (0--0.1, 0.1--0.2 ($\times$10$^{20}$ Mx), etc.),
are presented on a log$-$log scale in Figure \ref{fig8}. As we see in
the scatter plot, more than 95$\%$ of the magnetic elements have fluxes
less than 10$^{20}$ Mx, and patches with greater fluxes are more
scattered in size. In most cases, element sizes are slightly larger
for increasing elements fluxes. In order to find the relationship
between the size ($S$) and flux ($F$), which is expressed in terms
of arcsec$^2$ and 10$^{20}$Mx, respectively, we applied a linear fit
(blue line) on the mean values of each bin on a log$-$log scale.
It is found that the relationship between the size and flux  can be
expressed as $S\sim F ^{a}$, with a best-fit parameter of $a=0.69$
(Figure \ref{fig8}).

\subsection{Statistical properties of magnetic elements in the QS during 3 days}

In the next step, we cropped areas in HMI$/$SDO images (Figure \ref{fig1}C)
taken on 14$-$16 February 2011, and constructed a data cube with a cadence
of 45 s, which includes 5750 sequential frames. We are interested in finding
three consecutive days without a data gap in the quiet-Sun at solar disk center.
We aim to study the statistical properties of magnetic elements (\textit{e.g.},
lifetimes, and relationships between lifetime, flux and size) in the QS from
a large dataset. The total number of detected magnetic features is 185,150,
and a fraction of 22,526 features did survive for more than one frame, which
are labelled and tracked in time with the YAFTA's \textit{match features}
algorithm. This algorithm is working based on making masks, using overlapping
pixels in sequential images. The mean size of 22,526 events detected by
YAFTA follows a power law size distribution with exponent of -4.50, as
shown with a dashed line in a log$-$log scale in Figure \ref{fig9}.
To compare filling factors of magnetic elements that appear in the QS
(blue box in Figure \ref{fig1}A) with elements in the larger area
(red box in Figure \ref{fig1}A), which is included the other regions
than the QS, the filling factors of positive (red line) and negative
(blue line) polarities are plotted in Figure \ref{fig10}.
The \textit{Pearson} and \textit{Spearman} correlation coefficients
are computed and were found to be 0.34 and 0.33, respectively. The MLE
method is employed to find the power exponent of the lifetime-frequency
of patches, shown as a power law fit (dashed line in Figure \ref{fig11})
on a log$-$log scale. The exponent is found to be -1.85.
During these three days, one of the patches had a maximum lifetime of
876 minutes (see the electronic supplementary material (movie 1)),
and 11 patches were found to have lifetimes greater than 6 hours.
More than 90$\%$ of elements have lifetimes of less than 100 minutes.
The time series of the flux and size for four patches with long
duration lifetimes are shown in Figures \ref{fig12} and \ref{fig13},
respectively. The histogram of flux for patches appearing in the QS
is shown in Figure \ref{fig14}.
By ignoring the left hand-side of
the tail in the flux-frequency distribution, a linear fit with a slope
of -3.31 is found for the distribution of mean fluxes on a log$-$log
scale with more than 95$\%$ confidence. In this case, using the mean sizes
and fluxes over the lifetime of a feature instead of the \textit{instantaneous}
values can make a difference in the obtained exponents. The peak flux amounts to
4.5 $\times10^{17}$ Mx. In the QS, the maximum value for the flux belonging
to a patch is 1.66$\times10^{19}$ Mx.

The relationships between size, lifetime, and flux of patches in the QS
(indicated with $S$, $T$, and $F$, respectively), are shown in the
scatter plots as presented in Figures \ref{fig15}--\ref{fig17} on a
log$-$log scale. In the same manner a scatter plot is shown in
Figure \ref{fig8}, where the linear fits were applied to the mean
values obtained for each bin. The results are as follows: The
relation between mean size (arcsec$^2$) of magnetic elements during
their lifetimes and the lifetimes (minute) in the QS is defined by
$S\simeq T ^{c}$, where the fitting parameter $c$ is 0.25
(see the caption of Figure \ref{fig15}). The relation between
the mean flux (10$^{18}$ Mx) of magnetic elements during their
lifetimes and the lifetimes (minute) in the QS is $F\simeq T ^{e}$,
where the fitting parameter $e$ is 0.38
(see the caption of Figure \ref{fig16}). The relation between
mean size (arcsec$^2$) and mean flux (10$^{18}$ Mx) of magnetic
features during their lifetimes in the QS is $S\simeq F ^{g}$,
where the fitting parameter $g$ is 0.64 (see the caption of Figure \ref{fig17}).

As we can see in Figure \ref{fig15}, features with different
lifetimes have an excessive scatter in their sizes, but the
clustering of points shows that most of the elements with
smaller sizes (ranging from 2 to 8 arcsec$^2$) have lifetimes of
less than 100 minutes. In the same manner of Figure \ref{fig15},
in different ranges of patch lifetimes, there is an excessive
scatter in the flux (Figure \ref{fig16}). But, as it can be
seen in Figure \ref{fig17}, there is a clear relation between
the size and flux. The size gradually grows up with increasing
magnetic flux. The maximum value of size with 19.07 arcsec$^2$
belongs to a patch with a lifetime and flux of 171.5 minutes
and 10 $\times$10$^{18}$ Mx, respectively. The maximum value of
the flux with about 11$\times$10$^{18}$ Mx belongs to a patch
with a lifetime and size of 289 minutes and 18 arcsec$^2$,
respectively. The maximum value of lifetimes is 876 minutes,
belonging to a patch with a size and flux of 16.67 arcsec$^2$
and 8.74$\times$10$^{18}$ Mx, respectively.

Both the size and flux frequency distributions of the magnetic
features (\textit{i.e.,} Figures \ref{fig4}, and \ref{fig7})
represent the fluctuations in the tail of the right-hand side
of the histograms. It means that the statistical variations
are larger than the bin count numbers. Newmann (2005) explained
that this behavior is the characteristics of histograms
that follow a power law distribution with an exponent larger than 2.

The detrended fluctuation analysis (DFA) is used to investigate
the self-similarities in time series of new patches emerged in each sequence. In a time series,
self-similarity for events in different time steps ($Y_t$)
can be specified by the Hurst exponent ($H$).

\begin{equation} \label{Hurst}
Y_{at} = a^{-H} Y_t,
\end{equation}

where $a$ is a coefficient that determines any time later.
If $H$ takes the values in the ranges of [0,0.5] and [0.5,1],
we can say that the event has a long-term memory in its
anti-correlated and positively correlated behavior, respectively.
In the situation or $H$ = 0.5, there is an uncorrelated signal
(white noise) in a time series \cite{Mandel, Buld}.
For exponents greater than
unity the generalized Hurst exponent is introduced (for instance,
refer to Heneghan and McDarby (2000)).

The task of DFA, which is introduced by Peng
\textit{et al.} (1994), is fitting a line $(a_{j}x+b_j)$ for
each cumulative time series $C_{i,j}$. The cumulative time series is
defined as data series of components summation in each step over time.
If the time series with length $L$ is broken down into $M$ subwindows with
length $N$, the average of the mean square fluctuation for a whole
of time series is given as

\begin{equation} \label{DFA}
F(N) = \frac{1}{M}\sum_{j=1}^M \sqrt{ \left( \frac{1}{N}\sum_{i=1}^N
(C_{i,j} - a_{j}i - b_j)^2 \right)}.
\end{equation}

Using the value of all subwindows $F(N)$ versus $N$, frequently
expressed on a log$-$log scale, the Hurst exponent is evaluated \cite{Ali}.

The value of the Hurst exponent for the time series of all newly
emerged magnetic features during three days is found to be 0.82
(Figure \ref{fig18}). The value shows that there is a long-term
memory (auto-correlation) between different parts of the time series.
The systems with long-term correlations can be described by
small parts of the whole of system.

\section{Conclusions}\label{Conc}

The automatic method \textit{Yet Another Feature Tracking Algorithm}
(YAFTA) for segmentation and tracking of photospheric magnetogram
features taken by the \textit{Helioseismic and Magnetic Imager} (HMI)
was used to study statistical properties of magnetic elements.
A statistical study of elements during the year 2011 shows
a correlation coefficient of higher than 0.5 between filling factors
of positive and negative polarities. The slope values of
size-frequency distributions play an important role in small
magnetic elements on the solar surface. The flux-frequency
distribution follows a power law with a slope of about -2.15 $\pm$0.15,
closely corresponding to the value found by Parnell \textit{et al.} (2009).
A few percent of the difference in the result may be caused by the spatial resolution of instruments and the employed method for identifying flux concentrations.
The broken power law fits of both size and flux distributions
illustrate that probably two different regimes govern the evolution
of photospheric the magnetic elements. It is found that the relationship
between size ($S$) and flux ($F$) of magnetic elements can be expressed as

\begin{eqnarray} \label{relationships1}
S \propto F~^{0.69}.
\end{eqnarray}

Then, we focused on the statistical properties of patches in the QS
during three days with a cadence of 45 s. Among the patches that survived
more than one time during the analyzed sequence of images, more than 95\%
of the elements have lifetimes of less than 100 minutes. Moreover,
11 patches (about 0.05\% of elements) were found to have lifetimes more
than 6 hours and one of them lasted about 15 hours; the remaining
0.45\% have lifetimes ranging from 100 to 360 minutes. The power law
slopes are consistent with the following relationships between size ($S$),
flux ($F$) and lifetimes ($T$) of elements in the QS as follows:

\begin{eqnarray} \label{relationships2}
S \propto F~^{0.64}, ~~~~~ S \propto T~^{0.25}, ~~~~~ F \propto T~^{0.38}.
\end{eqnarray}

The exponents obtained from these relationships (between size $S$,
flux $F$, and lifetimes $T$) describe scaling laws between the observed
correlated quantities, as it is also evident from the Figures \ref{fig16}
and \ref{fig17}. According to Aschwanden (2015), deviations from
ideal power-law distributions can originate from three natural
effects, such as truncation effects, incomplete
sampling of the smallest events below some threshold, or contamination
by event-unrelated backgrounds. Since most sampled data are subject
to a detection threshold (of their sizes for instance), the
deviations from ideal power-laws can be explained by such data
truncation effects.

The power law exponents of most distributions of self-organized
criticality phenomena are consistent with power law-like distributions.
Suppose that the frequency distributions of two variables, $x$ and $y$,
follow the power-law function as $N(x) \sim x^{-\alpha_{x}} $ and
$N(y) \sim y^{-\alpha_{y}}$, one can directly derive the relationship
between the two variables $x$ and $y$. The correlation between $x$ and
$y$ can be expressed by $y \sim x^{\beta}$. If the value of $\beta$ approaches
unity, the variables $x$ and $y$ are proportional to each other.
The exponent $\beta$ can be obtained by performing a linear regression
fit to the logarithmic quantities $log(x)$ versus $log(y)$, as we
indicate in the scatter plots shown in Figures \ref{fig8}, \ref{fig15},
\ref{fig16}, and \ref{fig17}.

To test the consistency between the power law coefficients obtained
from the scatter plots (shown in relations \ref{relationships1} and
\ref{relationships2}), the following relation is used \cite{Aschwanden1}

\begin{eqnarray} \label{beta}
\beta = \frac{\alpha_{x}-1}{\alpha_{y}-1}.
\end{eqnarray}

\begin{figure} 
\centerline{\includegraphics[width=1.3\textwidth,clip=]{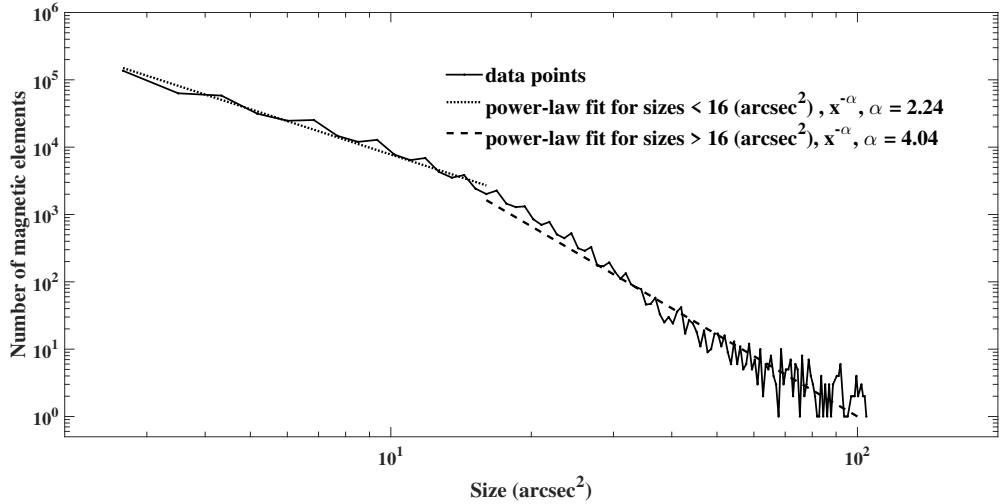}}
\caption[]{Size-frequency distributions of unsigned magnetic elements are drawn in log$-$log scale. A broken double linear fits representing power-law exponents is fitted. For sizes smaller and greater than 16 arcsec$^{2}$, the exponents were obtained -2.24 and -4.04, respectively. Running the MLE method on the size-frequency distribution represented that the exponent of power-law fit for the whole range of sizes is equal to -2.71.} \label{fig4}
\end{figure}

\begin{figure} 
\centerline{\includegraphics[width=1.3\textwidth,clip=]{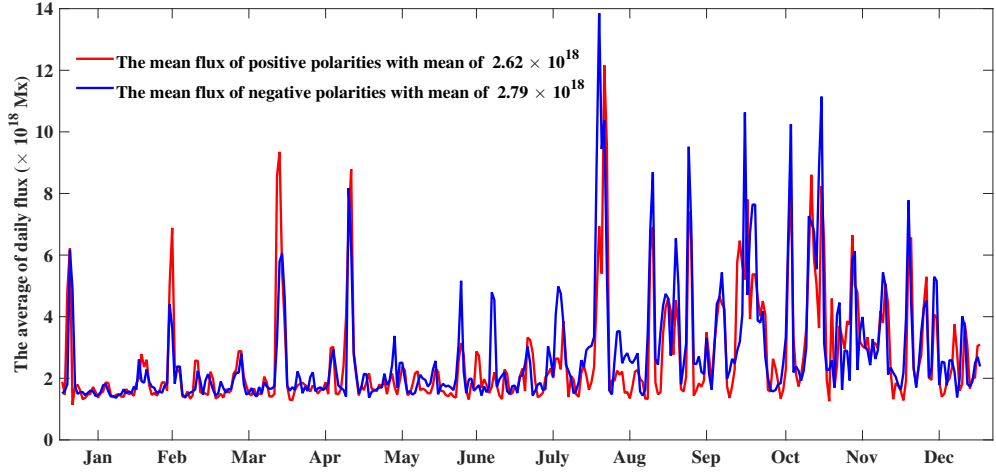}}
\caption[]{Daily average flux of positive (red) and negative (blue) elements are shown from January 1 until December 31, 2011.} \label{fig5}
\end{figure}

\begin{figure} 
\centerline{\includegraphics[width=1.3\textwidth,clip=]{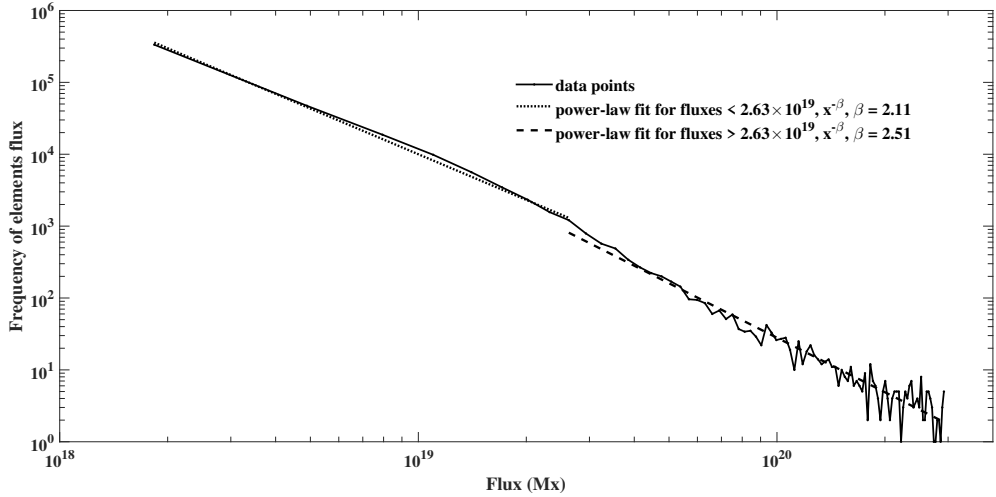}}
\caption[]{Histogram of polarities flux with their broken linear fits for fluxes smaller than 2.63$\times$10$^{19}$ Mx (the dotted line) and for fluxes greater than 2.63$\times$10$^{19}$ Mx(the dashed line) in log$-$log scale. The exponents of the two ranges are equal to -2.11 and -2.51, respectively. Exerting the MLE method on the whole range of flux frequencies gives the value of -2.15 for power-law exponent.} \label{fig7}
\end{figure}

\begin{figure} 
\centerline{\includegraphics[width=1.3\textwidth,clip=]{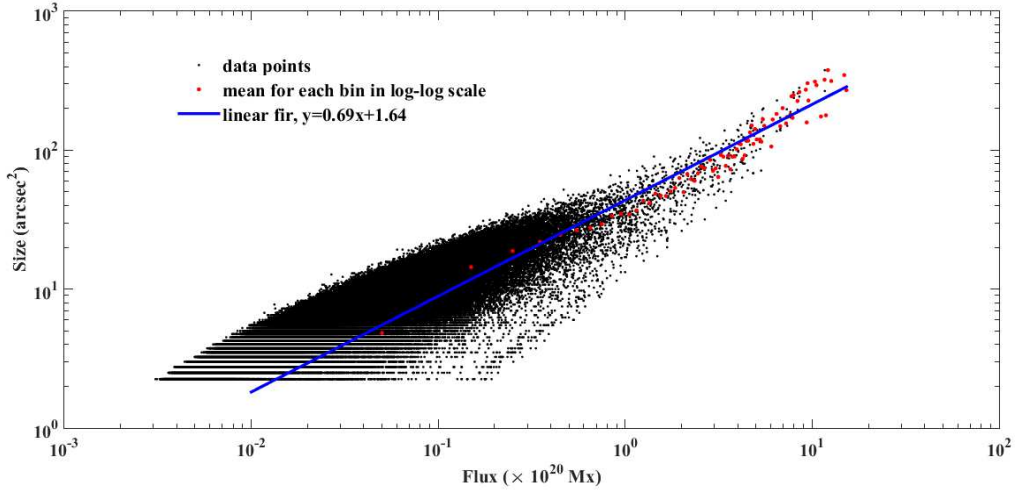}}
\caption[]{Relationship between the size of patches (arcsec$^2$) and their flux (Mx) in log$-$log scale.
The mean values for each bin (0 -– 0.1, 0.1 –- 0.2 $(\times 10^{20}$ Mx), etc.) for magnetic elements are shown as red points. The blue solid line displays the linear fit, $\log (S) = a\log (F) + b$ with $a$ = 0.69 $\pm$ 0.06 and $b$ = 1.64 $\pm$ 0.04 with 95$\%$ confidence level in fitting.} \label{fig8}
\end{figure}

\begin{figure} 
\centerline{\includegraphics[width=1.3\textwidth,clip=]{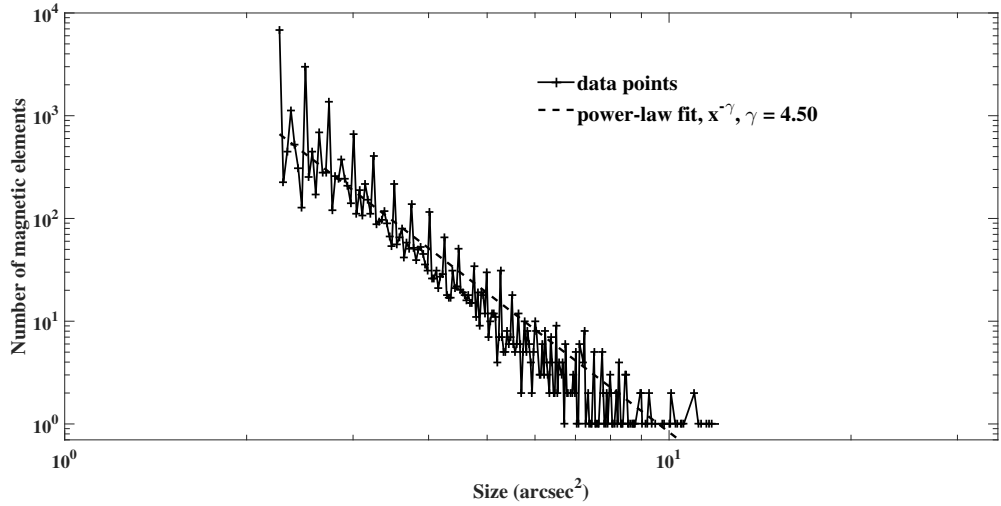}}
\caption[]{The distribution of mean size of the polarities during their life in the QS are plotted in log$-$log scale. According to the MLE method, the dashed line shows a power-law exponent of fitting with value of -4.50.} \label{fig9}
\end{figure}

\begin{figure} 
\centerline{\includegraphics[width=1.3\textwidth,clip=]{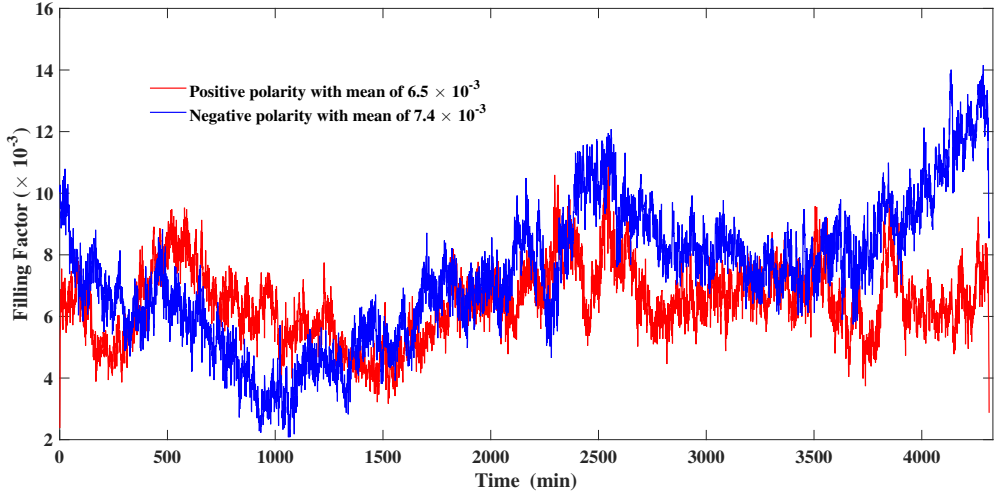}}
\caption[]{Filling factors (area) of positive (red) and negative (blue) polarities of magnetic elements are displayed for each minute during three days (14$-$16 February). The \textit{Pearson} and \textit{Spearman} correlations between these time series are about 0.34 and 0.33, respectively.} \label{fig10}
\end{figure}

\begin{figure} 
\centerline{\includegraphics[width=1.3\textwidth,clip=]{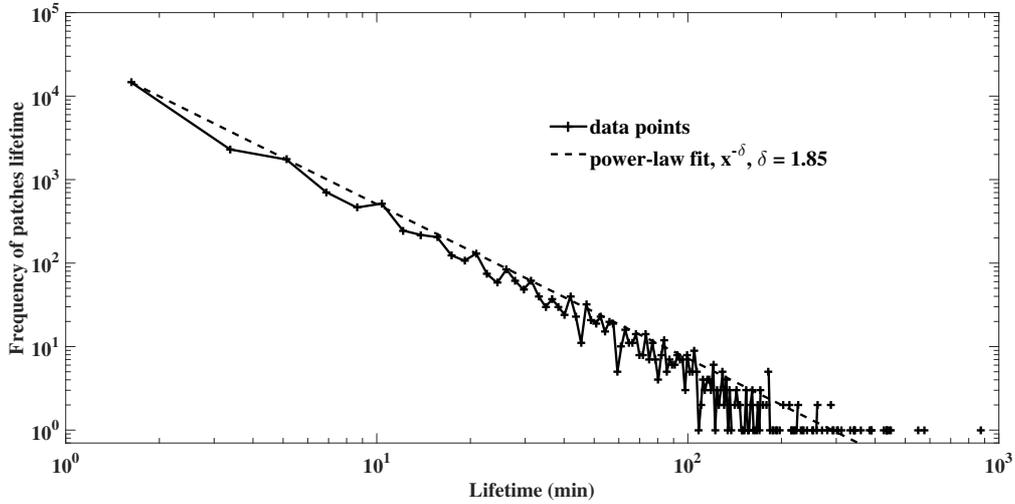}}
\caption[]{Histogram of patches lifetime and its linear fit (the dashed line) with the slope of -1.85 extracted from the MLE method. The maximum lifetime is obtained to be 876 minutes, and 11 patches were found to have lifetimes greater than 6 hours.} \label{fig11}
\end{figure}

\begin{figure} 
\centerline{\includegraphics[width=1.4\textwidth,clip=]{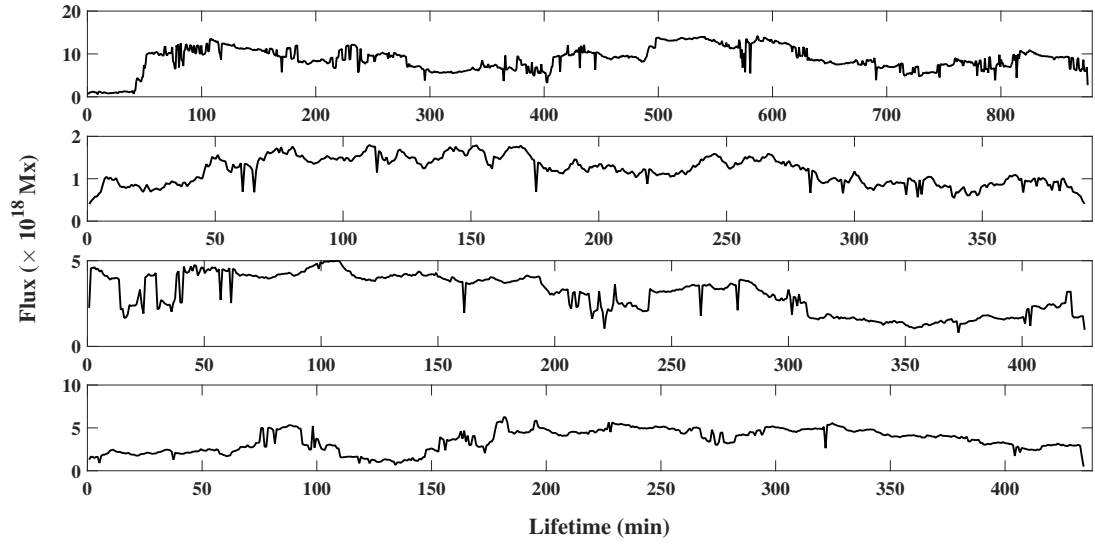}}
\caption[]{The flux variations of four patches with lifetimes more than 6 hours.} \label{fig12}
\end{figure}

\begin{figure} 
\centerline{\includegraphics[width=1.4\textwidth,clip=]{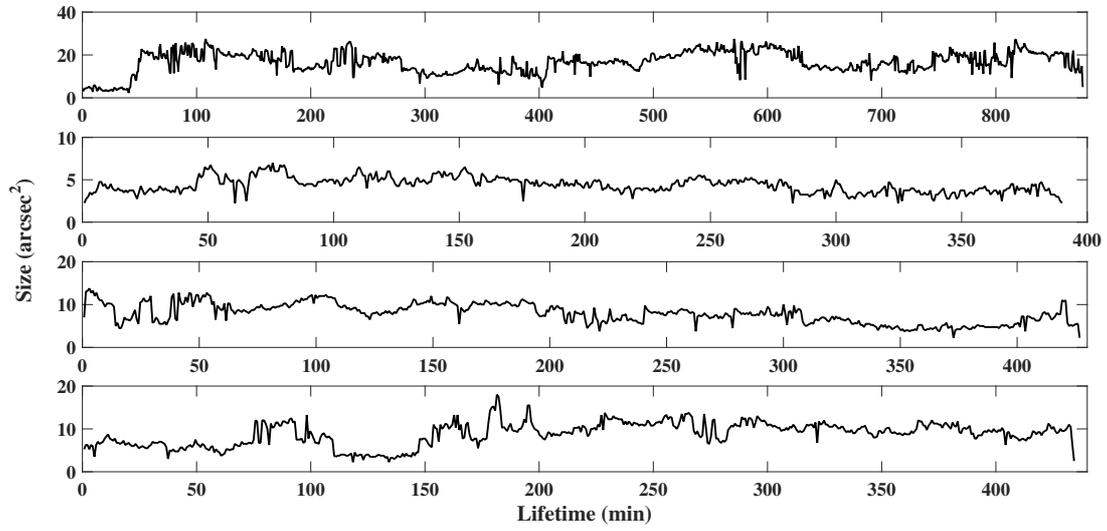}}
\caption[]{The size variations of the same four patches shown in figure \ref{fig12} during their lifetimes.} \label{fig13}
\end{figure}

\begin{figure} 
\centerline{\includegraphics[width=1.3\textwidth,clip=]{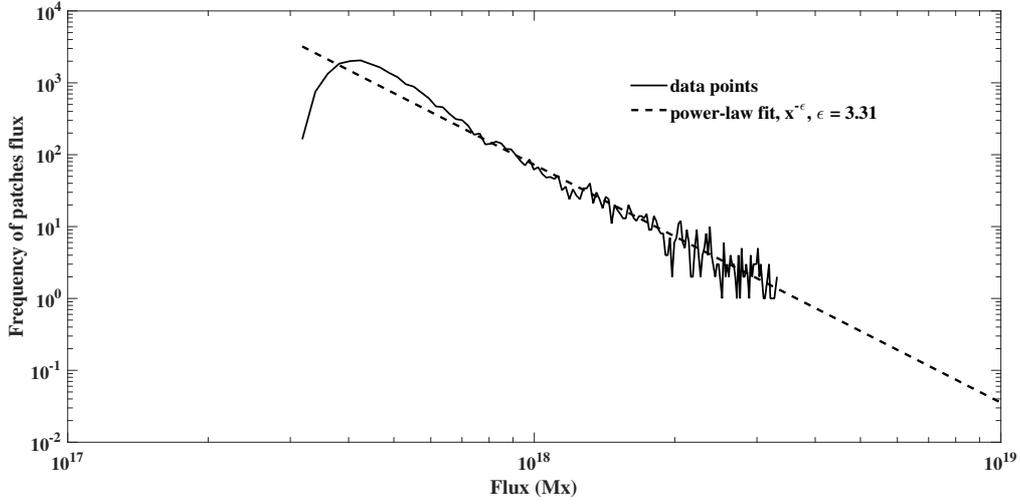}}
\caption[]{The distribution of mean flux of the magnetic elements during their life in the QS are plotted in log$-$log scale. According to the MLE method, the dashed line shows a power-law exponent of fitting with value of -3.31.
} \label{fig14}
\end{figure}

\begin{figure} 
\centerline{\includegraphics[width=1.3\textwidth,clip=]{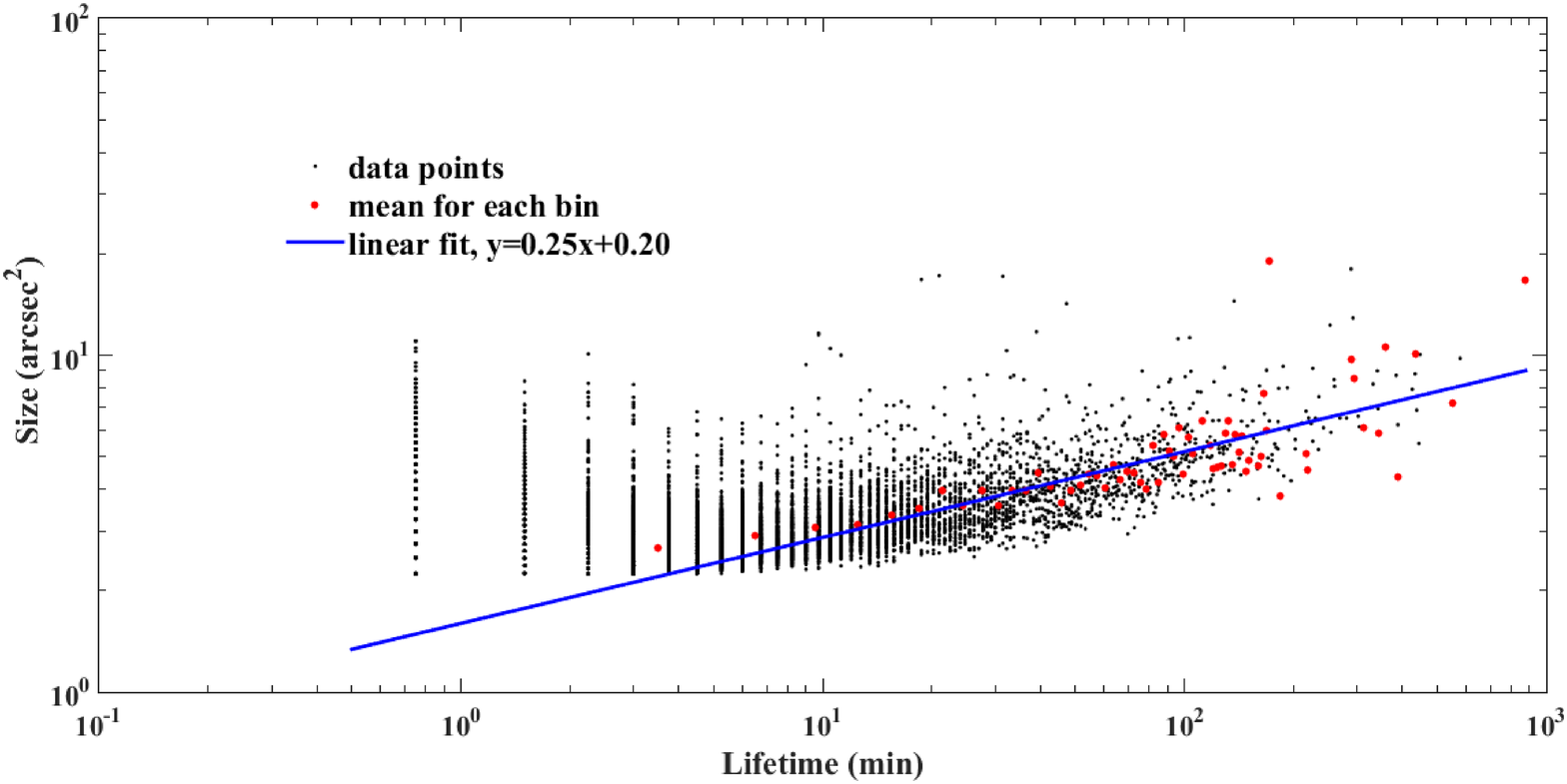}}
\caption[]{Relationship between mean size of the patches (arcsec$^2$) and their lifetime (min) appeared in the QS is shown in log$-$log scale. The mean values for each bin (0 -– 1, 1 –- 2 (min), etc.) for elements are indicated by red points. The blue solid line displays the linear fit, $\log (S) = c\log (T) + d$ with $c$ = 0.25 $\pm$ 0.06 and $d$ =  0.20 $\pm$ 0.1 with 95$\%$ confidence level in fitting.} \label{fig15}
\end{figure}

\begin{figure} 
\centerline{\includegraphics[width=1.3\textwidth,clip=]{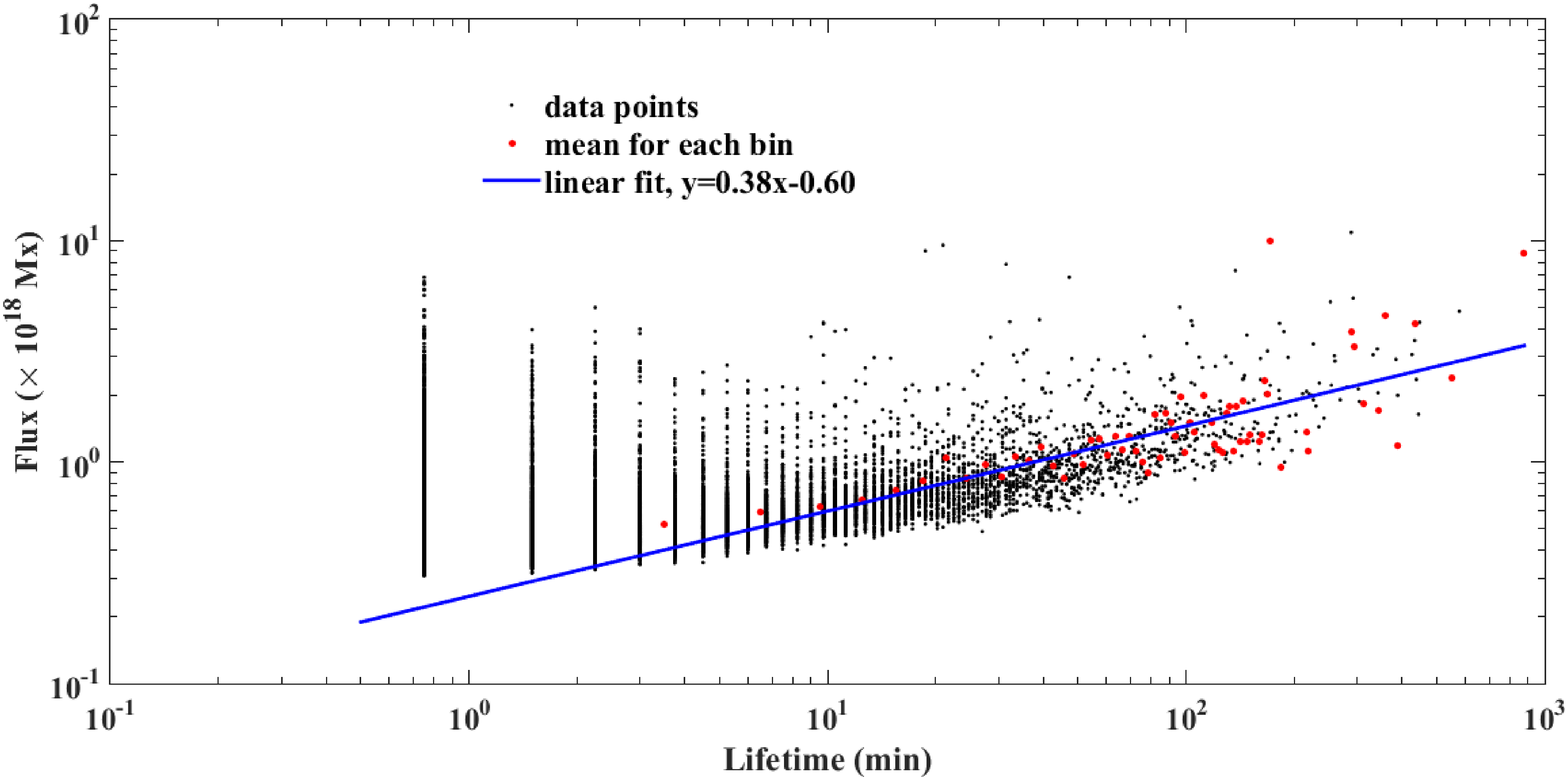}}
\caption[]{Relationship between mean flux of the patches (Mx) and their lifetime (min) appeared in the QS is shown in log$-$log scale. The mean values for each bin (0 -– 1, 1 –- 2 (min), etc.) for elements are indicated by red points. The blue solid line displays the linear fit, $\log (F) = e\log(T) + f$ with $e$ = 0.38 $\pm$ 0.09 and $f$ = -0.60 $\pm$ 0.17 with 95$\%$ confidence level in fitting.} \label{fig16}
\end{figure}

\begin{figure} 
\centerline{\includegraphics[width=1.3\textwidth,clip=]{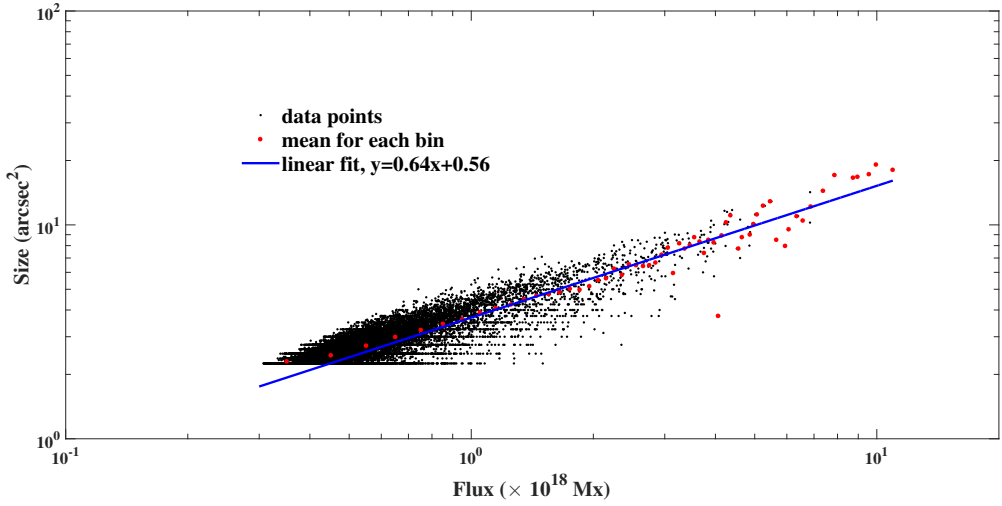}}
\caption[]{Relationship between mean flux of the patches (Mx) and their mean size (arcsec$^2$) during their lifetimes appeared in the QS is shown in log$-$log scale. The mean values for each bin (0 -– 0.1, 0.1 –- 0.2 ($\times 10^{18}$ Mx), etc.) for elements are indicated by red points. The blue solid line displays the linear fit, $\log (S) = g\log (F) + h$ with $g$ = 0.64 $\pm$ 0.06 and $h$ = 0.56 $\pm$ 0.04 with 95$\%$ confidence bounds.} \label{fig17}
\end{figure}

\begin{figure} 
\centerline{\includegraphics[width=1.5\textwidth,clip=]{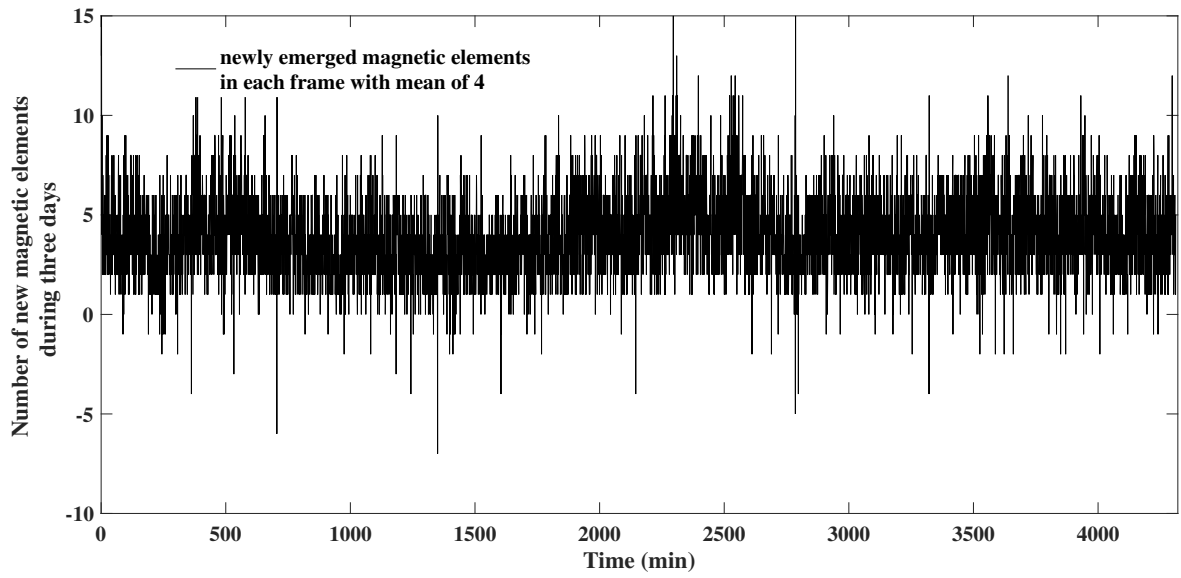}}
\caption[]{The time series of new magnetic elements appeared in each magnetogram for three-day data with a cadence of 45 s are shown. The average of newly emerged magnetic elements in each frame is equal to 4 elements. The value of Hurst exponent extracted by DFA method is equal to 0.82.}
\label{fig18}
\end{figure}

As it can be seen in Figures \ref{fig4} and \ref{fig7}, during the year 2011,
the power law indices $\alpha_{_{\small{F}}}$ and $\alpha_{_{\small{S}}}$,
which are extracted from histograms, were found to be 2.15 and 2.71,
respectively. According to Figures \ref{fig9}, \ref{fig11}, and \ref{fig14},
the extracted values (in the quiet Sun) for the power law indices
$\alpha_{_{\small{F}}}$, $\alpha_{_{\small{S}}}$, and $\alpha_{_{\small{T}}}$
are 3.31, 4.50, and 1.85, respectively. Using the power indices and
Equation (\ref{beta}), the corresponding correlations are as follows:

\begin{eqnarray} \label{relationships3}
S \propto F~^{0.67},~~~ S_{_{\rm\small{QS}}} \propto F_{_{\rm\small{QS}}}~^{0.66}, ~~~ S_{_{\rm\small{QS}}} \propto T_{_{\rm\small{QS}}}~^{0.24}, ~~~ F_{_{\rm\small{QS}}} \propto T_{_{\rm\small{QS}}}~^{0.36}.
\end{eqnarray}

It can be seen that these values are in agreement with values presented
by relations \ref{relationships1} and \ref{relationships2}.

The evaluation of the Hurst exponent for time series of newly emerged
magnetic elements on each day signifies that there is a correlation
between various parts of the signal, which corresponds to the case of
long-term memory. In other words, the self-similarity in the system
(called fractal dimensionality) occurs at a higher level. So, a
small fraction of the system can contain the information of the whole
system \cite{Aschwanden1}, which is typical for scale-free parameters.
It suggests that the system of solar magnetic features can be classified
as a self-organized criticality system, representing scale-invariance
characteristics of events with the ability of tuning itself as the system evolves.

In future work, we intend to extract all these statistical parameters
from active regions and coronal holes when they appear in the solar
equatorial region. Studying and comparing statistical properties of
patches appearing in the network and internetwork magnetic elements,
using both magnetograms and continuum intensity images taken by HMI
(for example refer to Yousefzadeh \textit{et al.}, 2016), will be the
focus of another project. Also, we are working on simulations of
time series of magnetic elements, to study the evolution of their flux
and size based on the obtained parameters and Monte-Carlo methods.
These simulated time series will then
be compared with observations using classifiers.

\ack The authors acknowledge the YAFTA group: C. E. DeForest, H. J. Hagenaar, D. A. Lamb, C. E. Parnell, and B. T. Welsch for making YAFTA results publicly available. 

\clearpage



\end{article}
\end{document}